\documentclass[aps,pre,twocolumn,showpacs,showkeys]{revtex4}

\usepackage{amsfonts}
\usepackage{amssymb}
\usepackage{psfig}
\usepackage{graphicx}

\begin{document}

\title{Quadratic solitons as nonlocal solitons}

\author{Ole Bang$^{1,3,4}$, Wies{\l}aw Z. Kr\'olikowski$^1$, 
        Nikola I. Nikolov$^{1,3,5}$, and Dragomir Neshev$^2$}

\affiliation{
$^1$Laser Physics Centre and $^2$Nonlinear Physics Group,
Research School of Physical Sciences \\ and Engineering, 
Australian National University, Canberra ACT 0200, Australia. \\
$^3$Informatics and Mathematical Modelling and $^4$Research Centre COM,
Technical University of Denmark, 2800 Kgs.~Lyngby, Denmark. \\
$^5$Ris{\o} National Laboratory, Optics and Fluid
Dynamics Department, OFD-128, P.O. Box 49, 4000 Roskilde, Denmark.}

\date{\today}

\begin{abstract}
We show that quadratic solitons are equivalent to solitons of a nonlocal 
Kerr medium.  This provides new physical insight into the properties of
quadratic solitons, often believed to be equivalent to solitons of an 
effective saturable Kerr medium. 
The nonlocal analogy also allows for novel analytical solutions
and the prediction of novel bound states of quadratic solitons. 
\end{abstract}



\maketitle

Quadratic nonlinear (or $\chi^{(2)}$) materials have a strong and fast 
electronic nonlinearity, which makes them excellent materials for the 
study of nonlinear effects, such as solitons \cite{SteHagTor96}. 
The main properties of quadratic solitons are well-known 
\cite{BurTraSkrTri02} and both 1+1 \cite{SchBaeSte96} and 2+1  
\cite{TorWanHagVanSteTorMen95} dimensional bright spatial solitons have 
been observed experimentally.
Unlike conventional solitons, which form due to a self-induced refractive
index change, the formation of quadratic solitons does not involve
any change of the refractive index.
Thus the underlying physics of quadratic solitons is often obscured by the
mathematical model.
Only recently Assanto and Stegeman used the cascading phase shift and
parametric gain to give an intuitive interpretation of effects, such as
self-focusing, defocusing, and soliton formation in $\chi^{(2)}$ materials
\cite{AssSte02}.

Nevertheless certain features of quadratic solitons, such as 
soliton interaction, are still without a physical interpretation.
Here we use the analogy between parametric interaction and nonlocality and
present a physically intuitive nonlocal theory, which is exact in predicting
the profiles of stationary quadratic solitons and which provides a simple
physical explanation for their properties including formation of bound
states.
The nonlocal analogy was applied recently by Shadrivov and Zharov to find 
approximate bright quadratic soliton solutions, but the nonlocal concept 
was not fully exploited to give a broad physical picture in the whole regime 
of excistence \cite{ShaZha02}.

We consider a fundamental wave (FW) and its second harmonic (SH) 
propagating along the $z$-direction in a quadratic nonlinear medium 
under conditions for type I phase-matching. 
The normalized
dynamical equations for the slowly varying envelopes $E_{1,2}(x,z)$ 
are then \cite{MenSchTor94-Ban97}
\begin{eqnarray}
   \label{phys1}
   & & i\partial_z E_1 + d_1\partial_x^2 E_1
       + E_1^*E_2 \exp(-i\beta z) = 0, \\
   \label{phys2}
   & & i\partial_z E_2 + d_2\partial_x^2 E_2
       + E_1^2 \exp(i\beta z) = 0.
\end{eqnarray}
In the spatial domain $d_1$$\approx$$2d_2$, $d_j$$>$0, and the coordinate 
$x$ represents a transverse spatial direction. The term $\partial_x^2 E_j$ 
then represents beam diffraction.
In the temporal domain $d_j$ is arbitrary and $x$ represents time. 
In this case $\partial_x^2 E_j$ represents pulse dispersion.
The parameter $\beta$ is the normalized phase-mismatch and $j$=1,2.

Physical insight into the properties of Eqs.~(\ref{phys1}-\ref{phys2}) 
may be obtained from the cascading limit, in which the phase-mismatch is 
large, $\beta^{-1}\rightarrow0$. 
Writing $E_2$=$e_2\exp(i\beta z)$ and assuming slow variation of 
$e_2(x,z)$ gives the nonlinear Schr\"odinger (NLS) equation
$i\partial_z E_1 + d_1\partial_x^2E_1 + \beta^{-1}|E_1|^2E_1 = 0$, 
in which the {\em local} Kerr nonlinearity is due to the coupling 
to the SH field $e_2$=$E_1^2/\beta$. 
The SH is thus slaved to the FW and the widths of the SH and FW are fixed.
The sign of the mismatch $\beta$ determines whether the effective Kerr 
nonlinearity is focusing or defocusing and thus the cascading limit 
predicts that both bright and dark quadratic solitons can exist.

However, even for stationary solutions the NLS equation is inaccurate,
since the term $\partial_x^2E_2$ is neglected. Thus it predicts that 
in higher dimensions bright solitons are unstable and will either 
spread out or collapse \cite{KivPel00}, whereas it is known that 
stable quadratic solitons exist in all dimensions and that collapse 
cannot occur in the $\chi^{(2)}$-system (\ref{phys1}-\ref{phys2}) 
\cite{BerMezRasWyl95,BerBanRasMez97}.
The stabilizing effect of the $\chi^{(2)}$ nonlinearity is often
described as being due to saturation of the effective Kerr nonlinearity
\cite{AssSte02,BerMezRasWyl95,WisTra02}.  
We show below that the nonlinearity is in fact nonlocal.

To obtain a more accurate model than that given by the cascading limit we 
assume a slow variation of the SH field $e_2(x,z)$ in the propagation 
direction only (i.e., only $\partial_z e_2$ is neglected).
The SH is still expressed in terms of the FW, but now the relation has 
the form of a convolution, leading to the {\em nonlocal equation for the FW}
\begin{eqnarray}
   \label{nonlocal_dynam}
   & & i\partial_z E_1 + d_1\partial_x^2E_1 + \beta^{-1} N(E_1^2)E_1^* = 0, \\
   \label{N}
   & & N(E_1^2) = \int_{-\infty}^\infty R(x-\xi)E_1^2(\xi,z) d\xi,
\end{eqnarray}
with $E_2$=$\beta^{-1}N\exp(i\beta z)$. 
Equations (\ref{nonlocal_dynam}-\ref{N}) clearly show that the interaction 
between the FW and SH is equivalent to the propagation of a FW in a medium 
with a nonlocal nonlinearity.
In the Fourier domain (denoted with tilde) the response function $R(x)$ 
is a Lorentzian $\widetilde{R}(k)$=$1/(1+s\sigma^2k^2)$, where 
$\sigma$=$|d_2/\beta|^{1/2}$ represents the degree of nonlocality and 
$s$=sign$(d_2\beta)$.
For $s$=+1, where the $\chi^{(2)}$-system (\ref{phys1}-\ref{phys2}) 
has a family of bright (for $d_1$$>$$0$) and dark (for $d_1$$<$$0$) 
soliton solutions \cite{BurKiv95_pla}, $\widetilde{R}(k)$ 
is positive definite and localized, giving
\begin{equation}
   R(x) = (2\sigma)^{-1}\exp(-|x|/\sigma).
\end{equation}
The cascading limit $\beta^{-1}\rightarrow 0$ is now seen to correspond to
the local limit $\sigma\rightarrow 0$, in which the response function
becomes a delta function, $R(x)\rightarrow\delta(x)$.

With the nonlocal analogy one can assign simple physically intuitive 
pictures to several important cases.
When the mismatch $|\beta|$ is large Eqs.~(\ref{nonlocal_dynam}-\ref{N}) 
reduce to the so-called weakly nonlocal equation with $\sigma\ll1$ 
\cite{KroBan01}. 
Similarly the nearly phase-matched limit when $\beta
\approx0$ corresponds to the strongly nonlocal limit with 
$\sigma\gg 1$, when Eqs.~(\ref{nonlocal_dynam}-\ref{N}) become 
effectively linear \cite{KroBanRasWyl01,BanKroWylRas02}.

For $s$=$-1$,  $\widetilde{R}(k)$ has poles on the real axis and the 
response function becomes oscillatory in nature with the Cauchy 
principal value $R(x)$=$(2\sigma)^{-1}\sin(|x|/\sigma)$. 
In this case the propagation of solitons has a close analogy with the 
evolution of a particle in a nonlinear oscillatory potential. In fact,  
it is possible to show that the oscillatory response function 
explains the fact that dark and bright quadratic solitons
radiate linear waves for $s$=$-1$ \cite{BurKiv95_pla}.  

It is important to notice that in contrast to the conventional nonlocal 
NLS equation describing, e.g., materials with a thermal \cite{LitMirFraYun75} 
or diffusion \cite{diffusion} type nonlinearity, liquid crystals
\cite{McLMurShe96-PecBrzAss02}, and photorefractive crystals 
\cite{MamZozMezAndSaf97}, the nonlocal response of the $\chi^{(2)}$
system depends on the square of the FW, not its intensity. 
Thus the phase of the FW enters into the picture and one cannot directly 
transfer the known nonlocal stability results for plane waves 
\cite{KroBanRasWyl01,WylKroBanRas02} and solitons 
\cite{KroBan01,BanKroWylRas02}. 

However, {\em for stationary fields the nonlocal system 
(\ref{nonlocal_dynam}-\ref{N}) represents an exact model for 
$\chi^{(2)}$ materials}. 
Therefore the known properties of nonlocal solitons in terms of profiles
directly apply to quadratic solitons.
Consider stationary solutions to Eqs.~(\ref{phys1}-\ref{phys2}) in the form
\begin{eqnarray}
  \hspace{-5mm}
  & & E_1(x,z)=a_1\phi_1(\tau)\exp(i\lambda z), \\
  & & E_2(x,z)=a_2\phi_2(\tau)\exp(i2\lambda z+i\beta z),
\end{eqnarray}
where the profile $\phi_j(\tau)$ is real, with $\tau=x|\lambda/d_1|^{1/2}$, 
$a_1^2=\lambda^2|d_2/(2d_1)|$, and $a_2$=$\lambda$. 
This scaling reduces the number of free parameters to one and
transforms Eqs.~(\ref{phys1}-\ref{phys2}) into the following system 
\cite{BurKiv95_pla}
\begin{equation}
  \label{chi2sol}
  \hspace{-12mm}
  s_1\phi_1'' -\phi_1 + \phi_1\phi_2=0, \;
  s_2\phi_2'' -\alpha\phi_2 + \phi_1^2/2=0,
\end{equation}
where $s_j$=sign$(\lambda d_j)$=$\pm1$, $\alpha$=$(2+\beta/\lambda)
|d_1/d_2|$, and prime denotes differentiation with respect to the
argument.
The properties of solitons described by Eqs.~(\ref{chi2sol}) are 
well-known \cite{BurKiv95_pla}. 
A family of bright (dark) solitons exist for $s_2$=$s_1$=$+1$
($s_2$=$-s_1$=$+1$) and $\alpha$$>$0. As discussed above we do not
consider the combinations $s_2$=$\pm s_1$=$-1$, 
for which solitons do not exist in the whole $\alpha$-space.

The second equation in Eqs.~(\ref{chi2sol}) has the formal solution
$\phi_2$=$\gamma N(\phi_1^2)$, with $\gamma$=$1/(2\alpha)$ and the 
nonlocal nonlinearity $N(\phi_1^2)$ given by Eq.~(\ref{N}). 
For sign$(s_2\alpha)$=$+1$ the response function is
$R(\tau)$=$(2\bar{\sigma})^{-1}\exp(-|\tau|/\bar{\sigma})$, with the 
degree of nonlocality $\bar{\sigma}$=$|\alpha|^{-1/2}$.
Inserting the SH into the first equation in Eqs.~(\ref{chi2sol}) then 
gives the {\em exact nonlocal model} for the FW in the $\chi^{(2)}$ 
system (\ref{chi2sol}) 
\begin{equation}
   \label{nonlocal_stat}
  \hspace{-8mm}
   s_1\partial_\tau^2\phi_1 -\phi_1 + \gamma \phi_1 
   \int_{-\infty}^{\infty}R(\tau-\xi)\phi_1^2(\xi)d\xi = 0,
\end{equation}
where $\gamma$ is the strength of the nonlocal nonlinearity.
Thus {\em $\chi^{(2)}$ solitons are equivalent to nonlocal solitons}.

In the weakly nonlocal case $\bar{\sigma}$$\ll$1 (i.e., $|\alpha|$$\gg$1) 
the response function $R(\tau)$ is much narrower than the FW intensity 
$\phi_1^2$. Taylor expanding $\phi_1^2$ under the 
integral in Eq.~(\ref{nonlocal_stat}) we then obtain the weakly 
nonlocal model \cite{KroBan01}
\begin{equation}
  \label{eq:weak}
  \hspace{-8mm}
  s_1\partial_\tau^2\phi_1 - \phi_1 + \gamma(\phi_1^2+
  \bar{\sigma}^2\partial_\tau^2\phi_1^2) \phi_1=0, 
\end{equation}
where $\phi_2$=$\gamma(1+\bar{\sigma}^2\partial_\tau^2)\phi_1^2$. 
This model has exact bright soliton solutions for 
$s_2$=$s_1$=$+1$ and $\alpha$$>$0 \cite{KroBan01}
\begin{equation}
   \label{sol:weak}
   \pm\tau = {\rm tanh}^{-1}(\rho) + 2\bar{\sigma}\tan^{-1}(2\bar{\sigma}\rho),
\end{equation}
where $\rho^2$=$(a_1^2-\phi_1^2)/(a_1^2+4\bar{\sigma}^2\phi_1^2)$,
$a_1^2$=$2/\gamma$ being the maximum intensity of the FW. 
Exact dark soliton solutions exist for $s_2$=$-s_1$=$+1$ with
nontrivial phase profiles \cite{KroBan01}.

For $|\alpha|$$\ll$1 the nonlocality is strong, $\bar{\sigma}$$\gg$1,
and we can expand the response function $R(\tau)$ in Eq.~(\ref{N}).
For bright solitons we then obtain the linear equation for the FW
\begin{equation}
  \label{eq:bright:strong}
  \hspace{-8mm}
  s_1\partial_\tau^2\phi_1 - \phi_1 + \gamma P_1R(\tau)\phi_1 = 0,
\end{equation}
where $\phi_2$=$\gamma P_1R(\tau)$. 
In this eigenvalue problem the FW power $P_1$=$\int_{-\infty}^\infty 
\phi_1^2(\tau)d\tau$ plays the role of the eigenvalue, and bright solitons
correspond to the fundamental mode of the waveguide structure created 
by the exponential response function.
For $s_2$=$s_1$=$+1$ and $\alpha>0$ Eq.~(\ref{eq:bright:strong}) has
exact bright soliton solutions in the form of the Bessel function of the 
first kind of order $2\bar{\sigma}$
\cite{Con73}
\begin{equation}
  \label{sol:bright:strong_odd}
  \hspace{-8mm}
  \phi_1(\tau) = A_1 J_{2\bar{\sigma}}[\bar{\sigma}^2\sqrt{2P_1R(\tau)}\,].
\end{equation}
For the single-soliton ground-state solution $P_1$ is found as the first
zero of the derivative, $J_{2\bar{\sigma}}^{\,\prime}(\sqrt{\bar{\sigma}^3
P_1})$=0, which assures that $\phi_1'(0)$=0. 
The amplitude $A_1$ is then found from the definition of $P_1$, giving
$A_1^2$$\approx$$P_1/2-[P_1/(\pi^2\bar{\sigma})]^{1/2}$.

\begin{figure}[h]
  \centerline{\scalebox{0.5}{\includegraphics{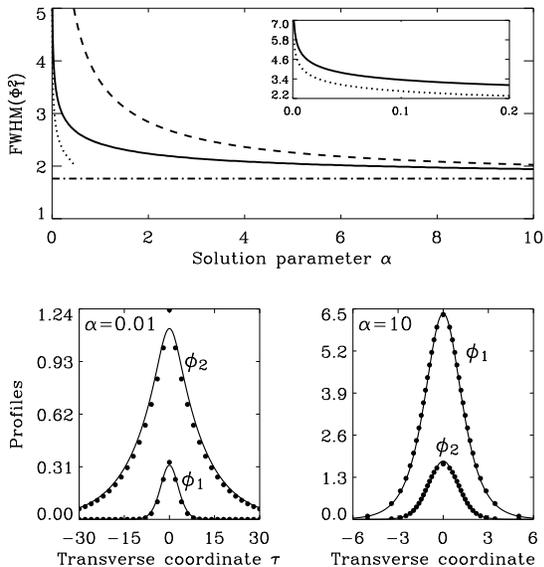}}}
  \caption{Top: Numerically found FWHM($\phi_1^2$) of bright quadratic 
  solitons versus 
  $\alpha$ (solid), and the weakly nonlocal (dashed), strongly nonlocal 
  (dotted), and cascading limit (chain-dashed) predictions.
  Bottom: Numerically found profiles (solid) and strongly nonlocal 
  (left: $\alpha$=0.01) and weakly nonlocal (right: $\alpha$=10) 
  solutions (dots). $s_2$=$s_1$=$+1$.
  \label{fig:bright}}
\end{figure}

In Fig.~\ref{fig:bright} we show the full width at half maximum of the FW 
intensity $\phi_1^2$ of bright quadratic solitons versus the phase-mismatch 
parameter $\alpha$.
The analytical solutions obtained using the nonlocal analogy correctly 
captures the increase of the soliton width with decreasing $\alpha$. 
The nonlocal model elegantly explains this effect:
Because of the convolution in the nonlinearity in Eq.~(\ref{nonlocal_stat}), 
representing a trapping potential or waveguide structure, this potential 
is always broader than the FW intensity profile itself, leading to its 
weaker confinement and larger width when the degree of nonlocality increases. 
The profiles shown in Fig.~\ref{fig:bright} further illustrates the excellent 
agreement of the numerical results and approximate nonlocal analytical 
solutions in both the weakly ($\alpha\gg1$) and strongly ($\alpha\ll1$) 
nonlocal limit.

\begin{figure}[h]
  \centerline{\scalebox{0.5}{\includegraphics{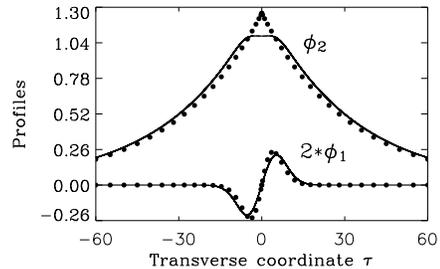}}}
  \caption{Numerically found bound state of two out-of-phase bright 
  solitons for $\alpha$=0.001 (solid) and the predicted strongly nonlocal 
  solution (\ref{sol:bright:strong_even}) (dashed). $s_2$=$s_1$=$+1$.}
  \label{fig:boundstate}
\end{figure}

The linear Eq.~(\ref{eq:bright:strong}) describing the strongly nonlocal 
limit further predicts the existence of multi-hump bright solitons.
Choosing $P_1$ as the $N$'th zero of the derivative, i.e., the $N$'th root 
in the equation $J_{2\bar{\sigma}}^{\, \prime}(\sqrt{\bar{\sigma}^3 P_1})$=0,
gives solitons with an odd number of humps ($2N-1$), as discussed in 
\cite{ShaZha02}.  
However, this does not exhaust all soliton solutions supported by 
the model (\ref{eq:bright:strong}).  
There also exist antisymmetric solitons with an even number of intensity 
peaks. 
If the power $P_1$ is found as the $N$'th zero of the Bessel function
itself, and not its derivative, i.e., as the $N$'th root in the equation
$J_{2\bar{\sigma}}(\sqrt{\bar{\sigma}^3P_1})$=0 (so that $\phi_1(0)$=0), 
then antisymmetric solitons with an even number of intensity peaks 
($2N$) exist with the form
\begin{equation}
  \label{sol:bright:strong_even}
  \hspace{-8mm}
  \phi_1(\tau) = s_\tau A_1 J_{2\bar{\sigma}}[\bar{\sigma}^2
                 \sqrt{2P_1R(\tau)}\,],
\end{equation}
where $s_\tau$=sign$(\tau)$.
When $P_1$, e.g., is fixed by the first zero of $J_{2\bar{\sigma}}$ then 
the solution (\ref{sol:bright:strong_even}) is a two-peak antisymmetric 
soliton, which can be interpreted as a bound state of two out-of-phase 
fundamental solitons. 
In Fig.~\ref{fig:boundstate} we have shown the bound state of two out-of-phase 
fundamental solitons predicted by Eq.~(\ref{sol:bright:strong_even}),
and the corresponding numerically found solution for 
$\alpha$=0.001.
We see again that the strongly nonlocal model provides an excellent
prediction of this novel bound state quadratic soliton solution.

In fact, all higher order solitons can be thought of as a bound 
state of a number of individual solitons.
Formation of such bound states follows naturally from the nonlocal
character of the nonlinear interaction. 
Consider two out-of-phase solitons, for which the intensity in 
the overlapping region is always zero. 
In local Kerr media the nonlinear change in the refractive index is 
decreased in the overlap region, as compared to the index change generated 
by a single soliton. This leads to a mutual repulsion of the solitons.
The nonlocality tends to increase the nonlinear change of the refractive 
index in the overlapping region, and for a sufficiently high degree of 
nonlocality, the index change may even be higher than for a soliton in 
isolation, despite the solitons being out-of-phase. 
This creates an attractive force and leads to formation of 
the bound state. 
 
In the strongly nonlocal limit the bright soliton is the {\em fundamental 
mode} of the waveguide structure $R(\tau)$ and much narrower than the 
waveguide. 
In contrast the dark soliton is a {\em first mode} of the waveguide $R(\tau)$
{\em at the cut-off} and as such its width is comparable with that of the 
waveguide. 
Hence the expansion procedure leading to Eq.~(\ref{eq:bright:strong}) is 
no longer justified. One can use it however, but for illustrative purposes 
only, to
show what type of dark soliton solutions can be expected in the strongly
nonlocal system. The linear equation for dark solitons in the strongly 
nonlocal limit will then have the form
\begin{equation}
  \label{eq:dark:strong}
  s_1\partial_\tau^2\phi_1 - \phi_1 + \gamma [A_1^2-Q_1R(\tau)]
  \phi_1 = 0,
  \hspace{-8mm}
\end{equation}
where $Q_1$=$\int_{-\infty}^\infty[A_1^2-\phi_1^2(\tau)]d\tau$ is the
complementary FW power and $\phi_2$=$\gamma[A_1^2-Q_1R(\tau)]$.
For $s_2$=$-s_1$=$+1$ and $\alpha>0$ Eq.~(\ref{eq:dark:strong}) has
exact dark soliton solutions in the form of the zero'th order Bessel
function
\begin{equation}
  \label{sol:dark:strong}
  \hspace{-8mm}
  \phi_1(\tau) = s_\tau \sqrt{2\alpha} J_0[\bar{\sigma}^2\sqrt{2Q_1R(\tau)}
  \,].
\end{equation}
For the fundamental single-soliton solution $Q_1$ is found as the 
first zero $J_0[\sqrt{\bar{\sigma}^3Q_1}\,]$=0, which 
gives $Q_1=5.8/\bar{\sigma}^{3}$ and assures that 
$\phi_1(0)$=0. As for bright solitons, choosing the $N$'th root gives 
novel multihump dark solitons with $2N-1$ dips in the intensity profile.
However, the background amplitude $A_1=2\alpha$ is fixed and does not
satisfy the self-consistency relation $Q_1$=$\int_{-\infty}^\infty
[A_1^2-\phi_1^2(\tau)]d\tau$. 

\begin{figure}[h]
  \centerline{\scalebox{0.5}{\includegraphics{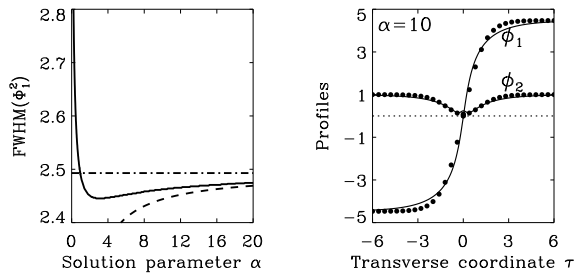}}}
  \caption{Left: Numerically found FWHM($\phi_1^2$) of dark solitons
  versus $\alpha$ (solid) and the weakly nonlocal (dashed) and cascading 
  limit (chain-dashed) predictions. Right: Numerically found profiles 
  (dots) and weakly nonlocal solutions (solid) for $\alpha$=10.  
  $\phi_1^2(\pm\infty)=2\alpha$, $\phi_2(\pm\infty)=1$, and $s_2$=$-s_1$=$+1$.}
  \label{fig:dark}
\end{figure}

In Fig.~\ref{fig:dark} we show the full width at half maximum of the FW 
intensity $\phi_1^2$ of dark quadratic solitons versus the mismatch 
parameter $\alpha$. The dark solitons have the constant background 
$\phi_1^2(\pm\infty)=2\alpha$, $\phi_2(\pm\infty)=1$. 
The analytical weakly nonlocal dark soliton solution exist for $\alpha>2$ 
and was taken from Ref.~\cite{KroBan01}.
Unlike bright solitons, whose width is a monotonic function of $\alpha$, 
dark solitons are seen to have a clear minimum width of the FW at 
$\alpha$=$\alpha_0\approx3.1$. 

Figure \ref{fig:dark} confirms that the weakly nonlocal model correctly 
predicts how the soliton width decreases for $\alpha>\alpha_0$ when the 
mismatch parameter decreases. 
This, as well as the appearance of the minimum in the 
soliton width, is again elegantly explained by the nonlocal analogy:
Because of the convolution in the nonlinearity in Eq.~(\ref{nonlocal_stat}), 
representing the trapping potential or waveguide structure, the contribution 
from the constant background tends to contract this potential. 
This leads to a stronger confinement and thus a smaller width of the 
soliton. However, this is only true as long as the amplitude of the 
trapping potential is not affected by nonlocality, as in the weakly 
nonlocal regime. For a high degree of nonlocality (i.e., smaller value 
of $\alpha$) not only the width of the trapping potential, but also its 
amplitude is affected. In this regime the nonlocality leads to a drop 
in the amplitude of the potential, resulting in a weaker confinement 
and an increase of the soliton width.
The profiles shown in Fig.~\ref{fig:dark} further illustrates the 
excellent agreement of the numerical results and approximate nonlocal 
analytical solutions in the weakly nonlocal limit $\alpha\gg1$.

In conclusion, we have used the analogy between parametric interaction 
in quadratic media and nonlocal Kerr-type nonlinearities to provide a
physically intuitive theory for quadratic solitons, which allows for a 
deeper physical insight into their properties and an exact description 
of their profiles. 
It is well-known that quadratic solitons may be characterized by a single 
solution parameter $\alpha$, which is an effective mismatch parameter
depending on both the real phase-mismatch and the power. 
We have shown that the nonlocal theory provides a simple and elegant 
explanation for how the soliton width depends on the mismatch 
$\alpha$.

In particular the nonlocal analogy provides simple physical models in 
both the large mismatch limit $\alpha$$\gg$1 and the near cut-off limit 
$|\alpha|$$\ll$1, corresponding to the regimes of weak and strong nonlocality,
respectively.
Our results show that the weakly nonlocal approximation gives an accurate 
description of quadratic solitons in a relatively broad range of their
existence domain. 
Also, the simple linear physics of the strongly nonlocal limit has enabled 
us to find novel bound states of bright quadratic solitons and explain their 
formation using the natural concept of the nonlocality-based attraction 
between out-of-phase constituent solitons. 

\begin{acknowledgments}
The research is supported by the Danish Technical Research Council
(Grant No.~26-00-0355) and the Australian Photonics Cooperative Research 
Centre. 
\end{acknowledgments}


\begin{thebibliography}{}
\bibitem{SteHagTor96}
   G. Stegeman, D.J. Hagan, and L. Torner,
   Opt. Quantum Electron. {\bf 28}, 1691 (1996).

\bibitem{BurTraSkrTri02}
   A.V.Buryak {\it et al.},
   Phys. Reports {\bf 370}, 63 (2002).

\bibitem{SchBaeSte96}
   R. Schiek, Y. Baek, and G.I. Stegeman,
   \pre {\bf 53}, 1138 (1996);

\bibitem{TorWanHagVanSteTorMen95}
   W.E. Torruellas {\it et al.},
   \prl {\bf 74}, 5036 (1995).

\bibitem{AssSte02}
   G. Assanto and G.I. Stegeman,
   Opt. Express {\bf 10}, 388 (2002).

\bibitem{ShaZha02}
   I.V. Shadrivov and A.A. Zharov,
   \josab {\bf 19}, 596 (2002).

\bibitem{MenSchTor94-Ban97}
   C.R. Menyuk, R. Schiek, and L. Torner,
   \josab {\bf 11}, 2434 (1994);
   O. Bang,
   {\it ibid.} {\bf 14}, 51 (1997).

\bibitem{KivPel00}
   Yu.S. Kivshar and D.E. Pelinovsky,
   Phys. Rep. {\bf 331}, 117 (2000).

\bibitem{BerMezRasWyl95}
   L. Berg\'e {\it et al.},
   \pra {\bf 52}, R28 (1995).

\bibitem{BerBanRasMez97}
   L. Berg\'e {\it et al.},
   \pre {\bf 55}, 3555 (1997).

\bibitem{WisTra02}
   F. Wise and P. Di Trapani,
   Optics \& Photonics News {\bf 13(2)}, 28 (2002).
  
\bibitem{BurKiv95_pla}
   A.V. Buryak and Yu.S. Kivshar, 
   Phys. Lett. A {\bf 197}, 407 (1995).

\bibitem{KroBan01}
   W. Krolikowski and O. Bang,
   \pre {\bf 63}, 016610 (2001).

\bibitem{KroBanRasWyl01}
   W. Krolikowski {\it et al.},
   \pre {\bf 64}, 016612 (2001).

\bibitem{BanKroWylRas02}
   O. Bang {\it et al.},
   \pre {\bf 66}, 046619 (2002).

\bibitem{WylKroBanRas02}
   J. Wyller {\it et al.},
   \pre {\bf 66}, 066615 (2002).

\bibitem{LitMirFraYun75}
   A.G. Litvak {\it et al.},
   Sov. J. Plasma Phys. {\bf 1}, 31 (1975).

\bibitem{diffusion}
   D. Suter and T. Blasberg, 
   \pra {\bf 48}, 4583 (1993).  

\bibitem{McLMurShe96-PecBrzAss02}
   D.W. McLaughlin, D.J. Muraki, and M.J. Shelly, 
   Physica D {\bf 97}, 471-497 (1996);
   M. Peccianti, K.A. Brzdakiewicz, and G. Assanto,
   \ol {\bf 27}, 1460 (2002).

\bibitem{MamZozMezAndSaf97}
   A. Mamaev {\it et al.},
   \pra {\bf 56}, R1110 (1997); 
   G.F. Calvo {\it et al.}, 
   Europhys. Lett. {\bf 60}, 847  (2002).

\bibitem{Con73}
   E.M. Conwell,
   \apl {\bf 23}, 328 (1973).
\end{thebibliography}
\end{document}